\documentclass{article}

\usepackage{amssymb}
\usepackage{fontenc}
\usepackage{times}
\usepackage{mathptmx}
\usepackage{color}
\usepackage{graphicx}
\usepackage{ifpdf}
\usepackage{psfrag}

\begin{document}

\title{Infrared Gluon Propagator from lattice QCD: results from large 
       asymmetric lattices}
\author{P. J. Silva\thanks{\textit{email}: psilva@teor.fis.uc.pt},
        O. Oliveira\thanks{\textit{email}: orlando@teor.fis.uc.pt} \\
        Centro de {F\'\i}sica Computacional \\
        Departamento de F{\'\i}sica \\
        Universidade de Coimbra \\
        3004-516 Coimbra \\
        Portugal }

\maketitle

\begin{abstract}
The infrared limit of the lattice Landau gauge gluon propagator is studied.
We show that the lattice data is compatible with the pure power law 
$( q^2 )^{2 \kappa}$ solution of the Dyson-Schwinger equations. Using various 
lattice volumes, the infinite volume limit for the exponent $\kappa$ is 
measured. Although, the results allow $\kappa = 0.498 - 0.525$, the lattice
data favours $\kappa \sim 0.52$, which would imply a vanishing zero momentum 
gluon propagator.
\end{abstract}

\section{Introduction and motivation}

The gluon propagator is a fundamental Green's function of Quantum 
Chromodynamics (QCD). Certainly, a good description of this two point
function is required to understand the non-perturbative regime of the theory. 
Moreover, in the Landau gauge, the gluon and the ghost propagators at zero 
momentum are connected with possible confinement mechanisms. In 
particular, the Kugo-Ojima confinement mechanism requires an infinite zero 
momentum ghost propagator \cite{KuOj79,Ku95} and the Zwanziger horizon 
condition \cite{Zw91,Zw92,Zw93} implies either reflection positivity violation
or a null zero momentum gluon propagator. 
From this last condition it can be proved
that the zero momentum ghost propagator should diverge \cite{Zw94}. 

The computation of the gluon and ghost propagators for the full range of
momenta cannot be done in perturbation theory. Presently, we have to rely 
either on Dyson-Schwinger equations (DSE) or on the lattice formulation of 
QCD; both are first principles approaches. In the first technique a 
truncation of an infinite tower of equations, together with the 
parametrizations of a number of vertices, namely the three gluon and the 
gluon-ghost vertices, is required to solve the equations. In the second, one 
has to care about finite volume and finite lattice spacing effects. 
Hopefully, the two 
solutions will be able to reproduce the same solution. 
In this article we are concerned with the lattice Landau gauge gluon 
propagator in momentum space,
\begin{equation}
  D^{ab}_{\mu\nu} (q) ~ = ~ \delta^{ab} \, 
           \left( \delta_{\mu\nu} \, - \, \frac{q_\mu q_\nu}{q^2} \right)
           \, \frac{ Z ( q^2 ) }{q^2} \, ,
\end{equation}
for pure gauge theory. The form factor $Z( q^2 )$ is the gluon dressing 
function.

Recently, in \cite{LeSm02} the DSE were solved for the gluon and ghost 
propagators in the deep infrared region. The solution assumes ghost dominance
and is a pure power law \cite{Review}, with the exponents of the two 
propagators related by a single number $\kappa$. For the gluon dressing 
function, the solution reads
\begin{equation}
  Z ( q^2 ) ~ =  ~ z \left( q^2 \right)^{2 \kappa} \, . \label{Zdse}
\end{equation}
The DSE equations allow a determination of the exponent $\kappa = 0.595$, 
which, for the zero momentum, implies a null (infinite) gluon (ghost) 
propagator, in agreement with 
the criteria described above, and a finite strong coupling constant defined 
from the gluon and ghost dressing functions \cite{FiAl02,FiAlRe02}. 
Renormalization group analysis based on the flow equation 
\cite{Pa04,LiPa04,FiGi04} were able to predict, for the solution (\ref{Zdse}),
the range of possible values for the exponent. The results of this analysis 
being $0.52 \, \le \, \kappa \, \le \, 0.595$, implying, again, a null 
(infinite) zero momentum gluon (ghost) propagator. A similar analysis of the 
DSE but using time-independent stochastic quantisation \cite{Zw02,Zw03} 
predicted the same behaviour and $\kappa = 0.52145$. 

The reader should be aware that not all solutions of the DSE predict a 
vanishing zero momentum gluon propagator. In \cite{AgNa03,AgNa04}, the authors
computed a solution of the DSE with a finite zero momentum gluon propagator. 
Moreover, in \cite{Bou05a,Bou05b} it was claimed that the gluon and ghost 
propagators in the deep infrared region are not connected via the same 
exponent $\kappa$ and that there isn't yet a clear theoretical answer 
concerning the value of zero momentum gluon propagator. According to the 
authors, the range of possible values goes from zero to infinity.

On the lattice, there are a number of studies concerned with the gluon 
propagator \cite{genglue0,genglue1,genglue2,genglue3,genglue4,genglue5}. 
On large lattices, the propagator was investigated in \cite{Lei99}. Although 
the lattices used had a limited access to the deep infrared region, the 
authors conclude in favour of a finite zero momentum gluon propagator, which 
would imply $\kappa = 0.5$ for the solution (\ref{Zdse}). In \cite{Bo01}, the 
infinite volume and continuum limits of the Landau gauge gluon propagator were
investigated using various lattices and, again, the data supported a finite 
zero momentum gluon propagator. 
In \cite{genglue2}, the three-dimensional lattice SU(2)
Landau gauge propagator was studied with the authors measuring an infrared
exponent compatible with the corresponding DSE solution. 

In order to access the deep infrared region, in 
\cite{NosSard,NosDub1,NosDub2} we have computed the gluon propagator with large
asymmetric lattices. The fits to (\ref{Zdse}) produced always $\kappa < 0.5$. 
However, the inclusion of corrections to the leading behaviour given by
(\ref{Zdse}) or when the gluon dressing function was modelled, either in the 
momenta region $q < 1$ GeV or for all the momenta, $\kappa$ becomes larger than
0.5. Moreover, in the preliminary finite volume study 
\cite{NosDub2,Fis05} of the two point gluon function it came out that 
$\kappa$ increases with the lattice volume and, in that sense, previously 
computed values should be regarded as lower bounds. The previous studies do 
not provide a clear answer about the zero momentum gluon propagator. 
In this paper, we update the results of our previous publications for the 
infrared region, using a larger set of lattices, that allows a better control 
of the infinite volume extrapolation. Two different extrapolations to the 
infinite volume are performed, namely the extrapolation of $\kappa$ and the
extrapolation of the propagator, giving essentially the same result. The main
conclusions from this investigation being that the lattice data is compatible
with the DSE solution (\ref{Zdse}) for the infrared gluon dressing function,
for momenta $\sim 150$ MeV or lower, and that the infinite volume extrapolation
of the lattice gluon propagator seems to favour
a vanishing zero momentum gluon propagator.

\section{Field Definitions and Notation}

In the lattice formulation of QCD, the gluon fields $A^a_\mu$ are replaced by 
the links
\begin{equation}
  U_\mu (x) ~ = ~ e^{ i a g_0 A_\mu (x + a \hat{e}_\mu / 2) }
  ~ + ~ \mathcal{O}( a^3 ) 
 ~ ~ ~ \in ~ SU(3)  \, ,
\end{equation}
where $\hat{e}_\mu$ are unit vectors along $\mu$ direction. QCD is a gauge 
theory, therefore the fields related by gauge transformations
\begin{equation}
  U_\mu (x) ~ \longrightarrow ~ g(x) ~ U_\mu (x) ~ 
 g^\dagger (x + a \hat{e}_\mu) \, ,
 \hspace{1cm} g \in SU(3) \, ,
\end{equation}
are physically equivalent. The set of links related by gauge transformations
to $U_\mu (x)$ is the gauge orbit of $U_\mu (x)$. 

The gluon field associated to a gauge configuration is given by
\begin{equation}
A_\mu (x + a \hat{e}_\mu / 2) ~ = ~
   \frac{1}{2 i g_0} \Big[ U_\mu (x)  -  U^\dagger_\mu (x) \Big] ~ - ~
   \frac{1}{6 i g_0} \mbox{Tr}
         \Big[ U_\mu (x)  -  U^\dagger_\mu (x) \Big] 
\end{equation}
up to corrections of order $a^2$.

On the lattice, due to the periodic boundary conditions, the discrete momenta
available are
\begin{equation}
  \hat{q}_\mu ~ = ~ \frac{2 \pi n_\mu}{a L_\mu} \, ,
  \hspace{0.5cm}  n_\mu ~ = ~ 0, \, 1, \, \dots \, L_\mu -1 \, ,
\end{equation}
where $L_\mu$ is the lattice length over direction $\mu$. The momentum
space link is
\begin{equation}
   U_\mu ( \hat{q} ) ~ = ~ \sum \limits_x ~ e^{-i \hat{q} x} ~
         U_\mu (x)
\end{equation}
and the momentum space gluon field
\begin{eqnarray}
   A_\mu ( \hat{q} ) & = & \sum \limits_x ~ 
         e^{-i \hat{q} \left(x + a \hat{e}_\mu / 2\right)} ~
         A_\mu (x + a \hat{e}_\mu / 2 ) \nonumber \\
         & = & \frac{e^{-i \hat{q}_\mu a/2}}{2 i g_0}
               \Bigg\{ ~
                   \Big[ U_\mu ( \hat{q} )  -  U^\dagger_\mu (- \hat{q}) \Big]
                   \, - \,
                   \frac{1}{3} \mbox{Tr}
                       \Big[ U_\mu ( \hat{q} )  -  U^\dagger_\mu (- \hat{q} )
 \Big] ~
               \Bigg\} ~ .
\end{eqnarray}

The gluon propagator is the gluon two point correlation function. The 
dimensionless lattice two point function is
\begin{equation}
 \langle A^a_\mu (\hat{q}) ~ A^b_\nu (\hat{q}')  \rangle ~  = ~
  D^{ab}_{\mu\nu} ( \hat{q} ) ~ V ~ \delta( \hat{q} + \hat{q}' ) ~ .  
\end{equation}
On the continuum, the momentum space propagator in the Landau gauge is
given by
\begin{equation}
D^{ab}_{\mu\nu} ( \hat{q} ) ~ = ~ \delta^{ab} ~
   \Big( \delta_{\mu\nu} ~ - ~ \frac{q_\mu q_\nu}{q^2} \Big) ~
   D( q^2 ) ~ . \label{propcont}
\end{equation}
Assuming that the deviations from the continuum are negligable, the lattice
scalar function $D( q^2 )$ can be computed directly from (\ref{propcont}) as 
follows
\begin{equation}
  D( q^2 ) ~ = ~ \frac{2}{(N^2_c-1)(N_d-1) V} \sum\limits_{\mu} ~ 
                 \langle ~ \mbox{Tr} \left[  A_\mu (\hat{q}) \, 
                                           A_\mu ( -\hat{q} ) \right] ~ \rangle
      \, ,
      \hspace{0.3cm} q \ne 0 \, ,
 \label{Dq2}
\end{equation}
and
\begin{equation}
  D( 0 ) ~ = ~     \frac{2}{(N^2_c-1) N_d V} \sum\limits_{\mu} ~ 
                 \langle ~ \mbox{Tr} \left[  A_\mu (\hat{q}) \, 
                                           A_\mu ( -\hat{q} ) \right] ~ \rangle  \, , \hspace{0.3cm}
       q = 0 \, ,
    \label{Dq20}
\end{equation}
where
\begin{equation}
  q_\mu \, = \, \frac{2}{a} ~ \sin \Big( \frac{\hat{q}_\mu a}{2} \Big) \, ,
\end{equation}
$N_c = 3$ is the dimension of the group, $N_d = 4$ the number of spacetime
dimensions and $V$ is the lattice volume.

\section{The Landau Gauge}

On the continuum, the Landau gauge is defined by
\begin{equation}
  \partial_\mu A_\mu ~ = ~ 0 \, .\label{landau_cont}
\end{equation}
This condition defines the hyperplane of transverse configurations
\begin{equation}
\Gamma ~ \equiv ~ \{A: ~ \partial \cdot A \, = \, 0 \} ~ .
\end{equation}
It is well known \cite{gribov} that $\Gamma$ includes more than one
configuration from each gauge orbit. In order to try to solve the
problem of the nonperturbative gauge fixing, Gribov suggested the use of
additional conditions, namely the restriction of physical configurational 
space to the region
\begin{equation}
   \Omega  ~ \equiv  ~ \{ A:~ \partial\cdot A \, = \, 0,~ 
                              \textit{M}[A] \, \geq \, 0 \} ~ \subset ~ \Gamma
 \, ,
\end{equation}
where $\textit{M}[A] ~ \equiv ~ - \nabla \cdot D[A] $ is the Faddeev-Popov
operator. However, $\Omega$ is not free of Gribov copies and does not provide 
a proper definition of physical configurations.

A suitable definition of the physical configurational space is given by the
fundamental modular region $\Lambda  \subset  \Omega$, the set of the absolute 
minima of the functional
\begin{equation}
   F_{A}[g] ~ =  ~ \int d^{4}x ~ \sum_{\mu} ~ 
    \mbox{Tr}\left[A_{\mu}^{g}(x)A_{\mu}^{g}(x)\right] \, .
 \label{fcont}
\end{equation}
In this article, the problem of Gribov copies\footnote{For the
$16^3 \times 128$ lattice, the gluon propagator was computed using both the 
gauge fixing method described here and the gauge fixing method described in
\cite{nosso}, which aims to find the absolute maxima of $F_U [g]$ (see 
equation (\ref{f})). For a similar number of configurations, i.e. 164 
configurations at $\beta = 6.0$, we found no clear differences in the 
propagators \cite{NosDub1}. Therefore, in this study it will be assumed that 
Gribov copies do not play a significant role. Note that in \cite{Zw03a}, 
it was shown that, in the continuum, the expectation values measured in 
$\Omega$ are free of Gribov copies effects.} 
on the computation of the gluon propagator will not
be discussed. For a numerical study on Gribov copies see \cite{SiOl04,Bog05} 
and references therein.

On the lattice, the situation is similar to the continuum theory. The Landau 
gauge is defined by maximising the functional
\begin{equation}
   F_{U}[g] ~ = ~ C_{F}\sum_{x,\mu} \, \mbox{Re} \, \{ \, \mbox{Tr} \,
       [ g(x)U_{\mu}(x)g^{\dagger }(x+\hat{\mu}) ] \, \}  \label{f}
\end{equation}
where 
\begin{equation}
   C_{F}  ~ = ~ \frac{1}{N_{d}N_{c}V}
\end{equation}
is a normalization constant.  Let $U_\mu$ be the configuration that maximises
$F_U[g]$ on a given gauge orbit. For configurations near $U_\mu$ on its gauge
orbit, we have
\begin{eqnarray}
 F_U [ 1 + i \omega (x)]  =  F_U [ 1 ]  +  
            \frac{C_F}{4} \sum_{x,\mu}  i \omega^a (x) 
            \mbox{Tr} \Big[ & &
                      \lambda^a \left( U_\mu (x) \, - \,
                                         U_\mu ( x - \hat{\mu}) \right)
                            ~ -  \nonumber \\
        & &
                            \lambda^a \left( U^\dagger_\mu (x) \, - \,
                                              U^\dagger_\mu ( x - \hat{\mu}) 
                                    \right)
\Big] \, ,
\end{eqnarray}
where $\lambda^a$ are the Gell-Mann matrices. By definition, $U_\mu$ is a 
stationary point of $F$, therefore 
\begin{eqnarray}
\frac{\partial F}{\partial \omega^a (x)} ~ = ~
\frac{i \, C_F}{4} \sum_{\mu} 
            \mbox{Tr} \Big[ & &
                      \lambda^a \left( U_\mu (x) \, - \,
                                         U_\mu ( x - \hat{\mu}) \right)
                            ~ -  \nonumber \\
        & &
                            \lambda^a \left( U^\dagger_\mu (x) \, - \,
                                              U^\dagger_\mu ( x - \hat{\mu}) 
                                    \right)
\Big] ~ = ~ 0 \, .
 \label{statf}
\end{eqnarray}
In terms of the gluon field, this condition reads
\begin{equation}
  \sum_{\mu} \mbox{Tr} \Big[ ~ \lambda^a 
                               \left( A_{\mu}(x + a \hat{\mu}/2) - 
                                      A_{\mu}(x - a \hat{\mu}/2)
                               \right) ~\Big] \, + \, 
   \mathcal{O} (a^2) ~ =  ~ 0 \, ,
\end{equation}
or
\begin{equation}
  \sum_{\mu} \partial_\mu A^a_{\mu}(x) ~ + ~
   \mathcal{O} (a) ~ =  ~ 0 \, ,
\end{equation}
i.e. (\ref{statf}) is the lattice equivalent of the continuum Landau gauge 
condition. The lattice Faddeev-Popov operator $M(U)$  is given by the second 
derivative of (\ref{f}).

Similarly to the continuum theory, on the lattice one defines the region of
stationary points  of (\ref{f})
\begin{equation}
   \Gamma ~ \equiv ~ \{U: ~ \partial \cdot A(U) \, = \, 0 \} \, ,
\end{equation}
the Gribov's region $\Omega$ of the maxima of (\ref{f}),
\begin{equation}
   \Omega ~ \equiv ~ \{U: ~ \partial \cdot A(U)=0 ~ \mbox{and} ~ M(U) \geq 0 \}
\end{equation}
and the fundamental modular region $\Lambda$ defined as the set of the 
absolute maxima of (\ref{f}). A proper definition of the lattice Landau gauge 
chooses from each gauge orbit, the configuration belonging to the interior of 
$\Lambda$.

In this work, the gauge fixing algorithm used is a Fourier accelerated 
steepest descent method (SD) as defined in \cite{davies}. In each iteration, 
the algorithm chooses
\begin{equation}
   g(x) \, = \, \exp \Bigg[ \hat{F}^{-1} \,
                               \frac{\alpha}{2} \,
                               \frac{ p_{max}^{2} a^{2}}{ p^{2} a^{2} } \,
                               \hat{F} \, 
           \left( \sum_\nu  \Delta_{- \nu} \left[ U_\nu ( x ) \, - \,
                                          U^\dagger_\nu ( x ) \right]
                            \, - \, \mbox{trace} \right) \Bigg]
\label{sd}
\end{equation}
where
\begin{equation}
\Delta_{-\nu} \left( U_\mu (x) \right) ~ =  ~ 
       U_\mu ( x - a \hat{e}_\nu) \, - \, U_\mu ( x ) \, ,
\label{delta}
\end{equation}
$p^{2}$ are the eigenvalues of $(-\partial^{2})$, $a$ is the lattice spacing
and $\hat{F}$ represents a fast Fourier transform (FFT). 
For numerical purposes, it is
enough to expand to first order the exponential in (\ref{sd}), followed by a 
reunitarization of $g(x)$. On the gauge fixing process, the quality of the 
gauge fixing is measured by
\begin{equation}
  \theta ~ = ~ \frac{1}{VN_{c}} ~ \sum_{x} \mbox{Tr}
      [\Delta(x)\Delta^{\dag}(x)] \label{theta}
\end{equation}
where 
\begin{equation}
 \Delta(x) ~ = ~ \sum_{\nu} \left[ U_\nu ( x - a \hat{e}_\nu) \, - \,
                                   U^\dagger_\nu (x) \, - \, \mbox{h.c.}
                                   \, - \, \mbox{trace} \right]
\end{equation}
is the lattice version of $\partial_\mu A_\mu \, = \, 0$.

\section{Lattice setup}

\begin{table}[t]
\begin{center}
\begin{tabular}{|r||@{\hspace{0.2cm}}r|
                    @{\hspace{0.2cm}}r|
                    @{\hspace{0.2cm}}r|
                    @{\hspace{0.2cm}}r|}
\hline
   Lattice           & Update
                     & therm.
                     & Sep.
                     & Conf \\
\hline
   $8^3 \times 256$   & 7OVR+4HB & 1500 & 1000 & 80  \\
   $10^3 \times 256$  & 7OVR+4HB & 1500 & 1000 & 80  \\
   $12^3 \times 256$  & 7OVR+4HB & 1500 & 1000 & 80  \\
   $14^3 \times 256$  & 7OVR+4HB & 3000 & 1000 & 80  \\
   $16^3 \times 256$  & 7OVR+4HB & 3000 & 1500 & 155 \\
   $18^3 \times 256$  & 7OVR+4HB & 3000 & 1500 & 40  \\
\hline
\end{tabular}
\caption{Lattice setup used in the study of the gluon 
propagator volume dependence.} \label{Uvol}
\end{center}
\end{table}

The gluon propagator was computed using the SU(3) pure gauge, Wilson action,
$\beta = 6.0$ configurations for the lattices reported in table \ref{Uvol}. 
All configurations were generated with the MILC code 
{\tt http://physics.indiana.edu/\~{ }sg/milc.html}.
The table describes the combined overrelaxed+heat bath Monte Carlo sweeps, the
number of thermalisation sweeps of the combined overrelaxed+heat bath steps,
the number of combined sweeps separating each configuration
and the total number of configurations for each lattice. The statistical
errors on the propagators were computed using the jackknife method.

\begin{center}
\begin{figure}[t]
\begin{minipage}[b]{\textwidth}
\centering
\includegraphics[scale=0.5]{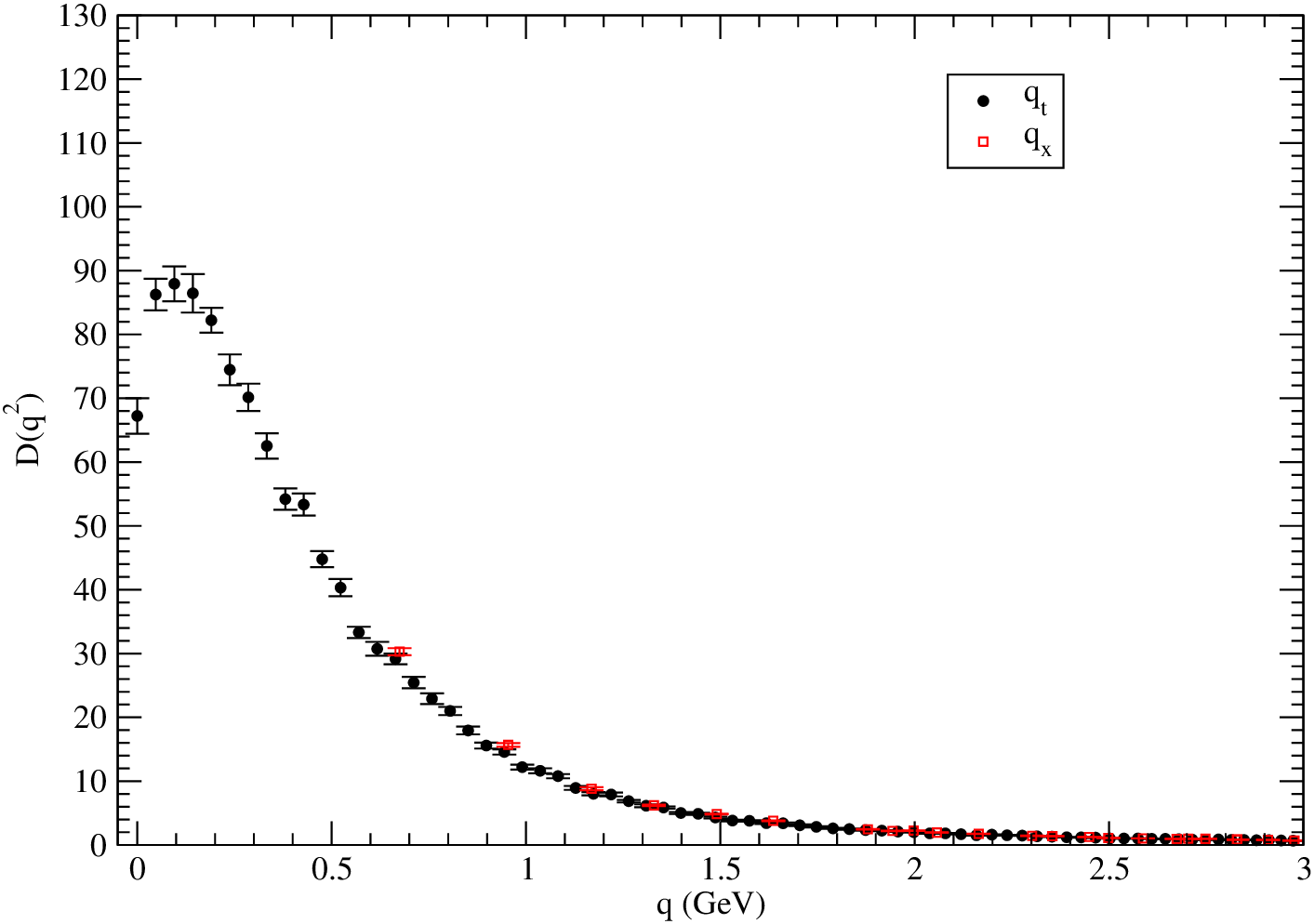}
\end{minipage}

\vspace{1.1cm}

\begin{minipage}[b]{\textwidth}
\centering
\includegraphics[scale=0.5]{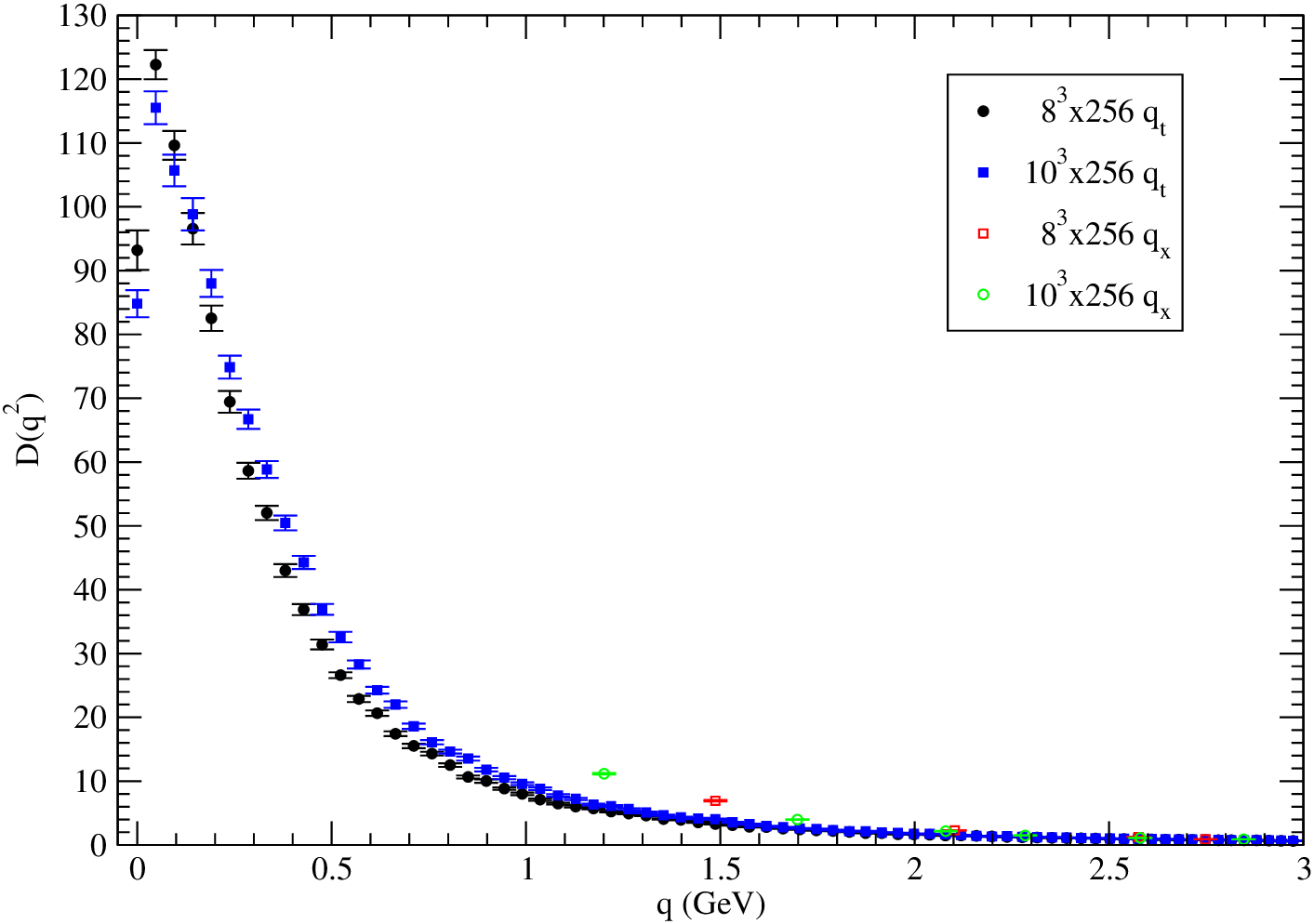}
\end{minipage}
\caption{Bare gluon dressing function for all momenta for the largest (top)
and the two smallest (bottom) lattices. 
The figure shows the gluon propagator computed
with pure temporal momenta $q_t$ and pure spatial momenta $q_x$. For the 
spatial momenta a $Z_3$ average was performed.}
\label{Zall}
\end{figure}
\end{center}

\begin{center}
\begin{figure}[t]
\begin{minipage}[b]{\textwidth}
\centering
\includegraphics[height=12cm,width=\textwidth]{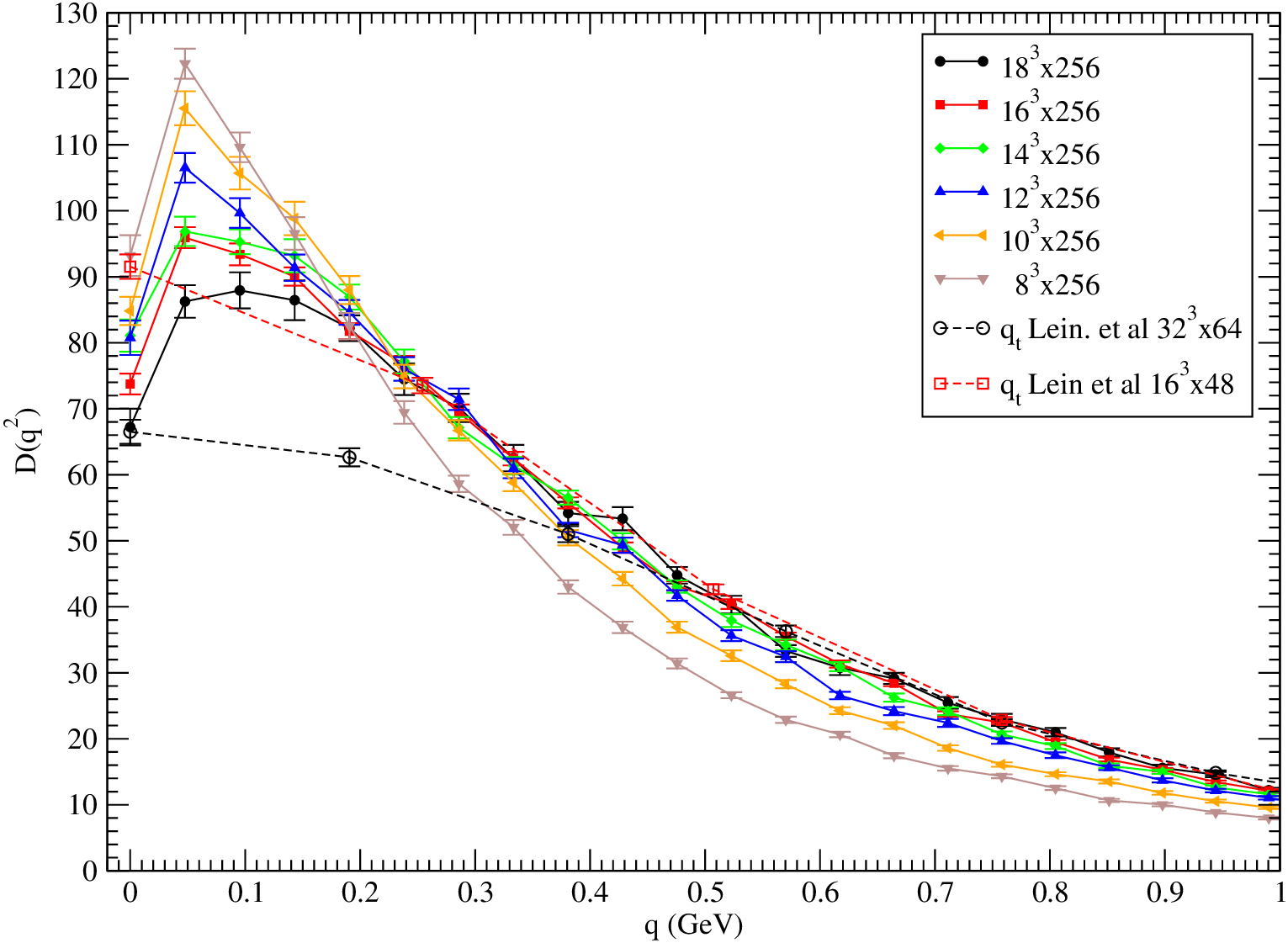}
\end{minipage}
\caption{Bare gluon dressing function for temporal momenta for all lattices
volumes. For comparisation, the figure includes the $16^3 \times 48$ and 
$32^3 \times 64$ propagators computed at $\beta = 6.0$ in \cite{Lei99}.}
\label{ZallLein}
\end{figure}
\end{center}

The propagators computed in this study have finite volume effects 
\cite{NosDub2}. An example is seen in figure \ref{Zall} where the bare gluon 
propagator, $D(q^2) = Z(q^2)/q^2$, is plotted for temporal momenta $q_t$ and 
spatial momenta $q_x$ for the smallest and largest lattices. 
Figure \ref{ZallLein} shows $D(q^2)$ for temporal $q \le 1$ GeV for the 
volumes considered in this study, including the two lattices
$16^3 \times 48$ and $32^3 \times 64$, $\beta = 6.0$, temporal momenta 
propagators from \cite{Lei99}.
For the lattices
discussed here, it was observed that the difference between the propagators
computed using only pure temporal or pure spatial momenta becames smaller as 
the spatial extension of the lattice increases. 
Moreover, the differences vanish
for sufficient high momenta and become larger for smaller momenta. 
Furthermore, it was observed that, for the smallest momenta, the propagator 
decreases as the lattice volume increases, in agreement with what was observed
in previous studies of the volume dependence \cite{Bo01,NosDub2}. In this 
work, we will not discuss what should be the right choice of momenta to 
minimize finite volume effects. Since our smallest momenta are pure temporal 
momenta, the results of the various lattices will be combined to investigate 
the volume dependence of the infrared gluon propagator. In the following and 
in order to access the infrared region, only the pure temporal momenta will be
considered.

\section{The Infrared Gluon Dressing Function}

\begin{table}[t]
\begin{center}
\begin{tabular}{|r||@{\hspace{0.3cm}}r@{\hspace{0.3cm}}r|
                                     r@{\hspace{0.3cm}}r|
                                     r@{\hspace{0.3cm}}r|}
\hline
   Lattice     & \multicolumn{2}{c|}{$z (q^2)^{2 \kappa}$} 
               & \multicolumn{2}{c|}{$z (q^2)^{2 \kappa} ( 1 + a q^2 )$} 
               & \multicolumn{2}{c|}{$z (q^2)^{2 \kappa} ( 1 + a q^2 + b q^4)$} \\
\hline
                      &                      &     & & & & \\
   $ 8^3 \times 256$  & $0.4496^{+22}_{-29}$   & 2.14 &
                        $0.4773^{+37}_{-52}$   & 0.02 &
                        $0.4827^{+75}_{-74}$   & 0.00 \\
                      &                      &     & & & & \\
   $10^3 \times 256$  & $0.4650^{+31}_{-37}$   & 0.10 &
                        $0.4827^{+49}_{-68}$   & 0.25 &
                        $0.4765^{+104}_{-99}$  & 0.14 \\
                      &                      &     & & & & \\
   $12^3 \times 256$  & $0.4663^{+30}_{-36}$   & 1.19 &
                        $0.4822^{+51}_{-69}$   & 0.21 &
                        $0.4849^{+94}_{-97}$   & 0.18 \\
                      &                      &     & & & & \\
   $14^3 \times 256$  & $0.4918^{+26}_{-40}$   & 0.09 &
                        $0.5053^{+52}_{-67}$   & 0.16 &
                        $0.4992^{+100}_{-80}$   & 0.06 \\
                      &                      &     & & & & \\
   $16^3 \times 256$  & $0.4859^{+22}_{-24}$   & 0.40 & 
                        $0.5070^{+36}_{-50}$   & 0.44 & 
                        $0.5131^{+67}_{-64}$   & 1.03  \\
                      &                      &     & & & & \\
   $18^3 \times 256$  & $0.5017^{+49}_{-40}$   & 0.20 & 
                        $0.5169^{+89}_{-70}$   & 0.00 &
                        $0.514^{+12}_{-15}$    & 0.00     \\
                      &                      &     & & & & \\
\hline
\end{tabular}
\caption{$\kappa$ and $\chi^2/d.o.f.$ from fitting the different lattices. 
In all fits, the range of momenta considered starts with the lowest 
non-vanishing momentum. The errors shown are statistical and were computed 
using the bootstrap method, with the number of bootstrap samples being about 
ten times the number of configurations. 
For the momenta range considered in the fits, the
correction associated with $a q^2$ and $a q^2 + b q^4$ to the pure power law
are below 20\% and 30\%, respectively. For the smallest lattice, the 
corrections are larger. However, this lattice is never used in the 
extrapolations.}
\label{kappa}
\end{center}
\end{table}

\begin{figure}[t]
\begin{minipage}[b]{\textwidth}
\centering
\includegraphics[origin=c,scale=0.5]{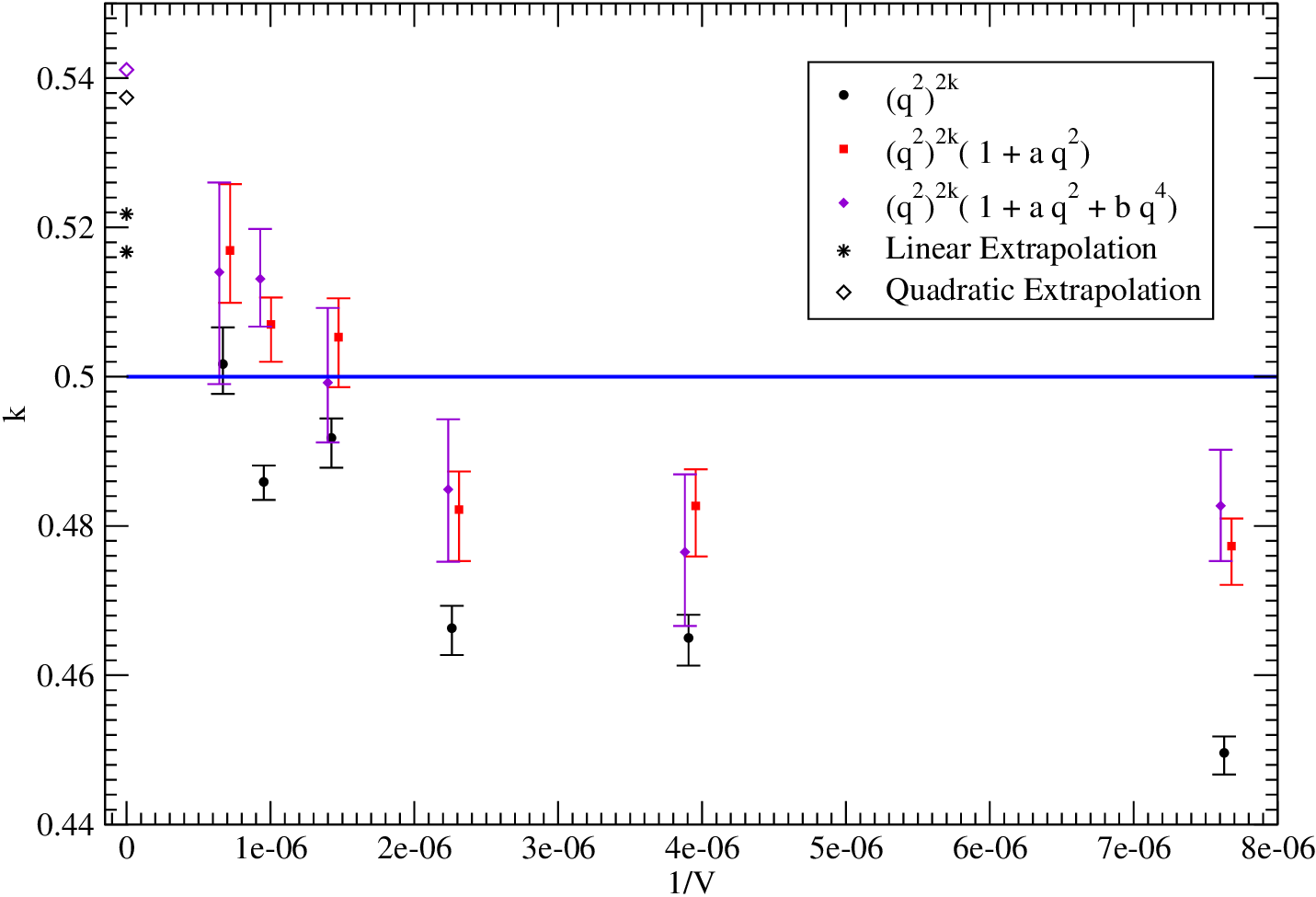}
\end{minipage} 
\caption{$\kappa$ as a function of the inverse volume. Note that in order to 
make the plot more clear, the $\kappa$ from the fits to the corrections to the
pure power law were shifted in the $1/V$ axis.}
\label{kappaV}
\end{figure}

An analytical solution of the DSE for the gluon dressing function is given by 
(\ref{Zdse}). For $\beta = 6.0$, the lattice spacing 
is $a^{-1} = 1.94(5)$ GeV \cite{LatSpa},
therefore our smallest nonzero momentum is $47.6 \pm 1.2$ MeV. 
Since we do not know at
what momenta the above solution sets in, besides the pure power law, we will
consider also polynomial corrections to the leading behaviour, i.e. 
\begin{equation}
   Z(q^2 ) ~ = ~ z \,  \left( q^2 \right)^{2 \kappa} \, 
                 \left\{  1 ~ + \sum_{n} a_n \left( q^2 \right)^n \right\}
 \, ,
\end{equation}
with $n = 1$, $2$. In order to be as close as possible to the infrared region,
in all fits only the smallest range of momenta will be considered, i.e. all 
reported fits have one degree of freedom.  
The $\kappa$ and $\chi^2/d.o.f.$ for the various fits are given in table 
\ref{kappa}. In general, the data is well 
described by any of the above functions. The exception being the smallest 
lattice for a pure power law. $\kappa$ shows finite volume effects, becoming 
larger for larger volumes. Note that for the largest lattices, the various
$\kappa$ computed with the corrections to the pure power law agree within one 
standard deviation. Moreover, for the largest lattice, the various $\kappa$ 
agree with each other within less than 1.5 standard deviations.
Figure \ref{kappaV} shows the results of the fits to $\kappa$ as function of 
the inverse of the lattice volume.

The infinite volume $\kappa$ can be estimated combining the results of table 
\ref{kappa}. Assuming, for each of the reported fits to the dressing 
function, either a linear or a quadratic dependence on $1/V$ for $\kappa (V)$
and excluding the smallest lattice, it comes out that the figures on the first
column, the pure power law, are not described by any of these functional
forms\footnote{The smallest $\chi^2/d.o.f.$ being larger than 5. This applies
even if one considers a cubic fitting function.}.
Figure \ref{kappaV} includes, besides the $\kappa$ values for each volume,
the extrapolated $\kappa$ from fitting $\kappa (1/V)$ to a linear or quadratic 
function of $1/V$. Note that the points are computed independently for each
of the corrections to the pure power law considered. The extrapolated $\kappa$
agree within one standard deviation, with the quadratic fits to $\kappa (1/V)$ 
giving larger errors and lying above the linear fits. For the linear 
extrapolation the results give
\begin{equation}
  \kappa_{\infty} ~ = ~ 0.5167 (55) \, ,
  \hspace{0.7cm}
  \kappa_{\infty} ~ = ~ 0.5218 (80) \label{kinfty}
\end{equation}
when the extrapolation uses the data from the quadratic or the quartic
corrections to (\ref{Zdse}). The corresponding $\chi^2/d.o.f.$ being 1.44 and
0.49. For the quadratic fit the extrapolated values are
\begin{equation}
  \kappa_{\infty} ~ = ~ 0.537 (13) \, ,
  \hspace{0.7cm}
  \kappa_{\infty} ~ = ~ 0.541 (20) \, ;\label{kinfty2}
\end{equation}
the $\chi^2/d.o.f.$ being 0.72 and 0.10, respectively. 
The errors $\kappa$ were computed assuming gaussian error propagation. 
The results of the extrapolation to the infinite lattice volume, point to a
value of $\kappa$ in the range $0.51-0.56$.

\begin{figure}[t]
\begin{minipage}[b]{\textwidth}
\centering
\includegraphics[origin=c,scale=0.5]{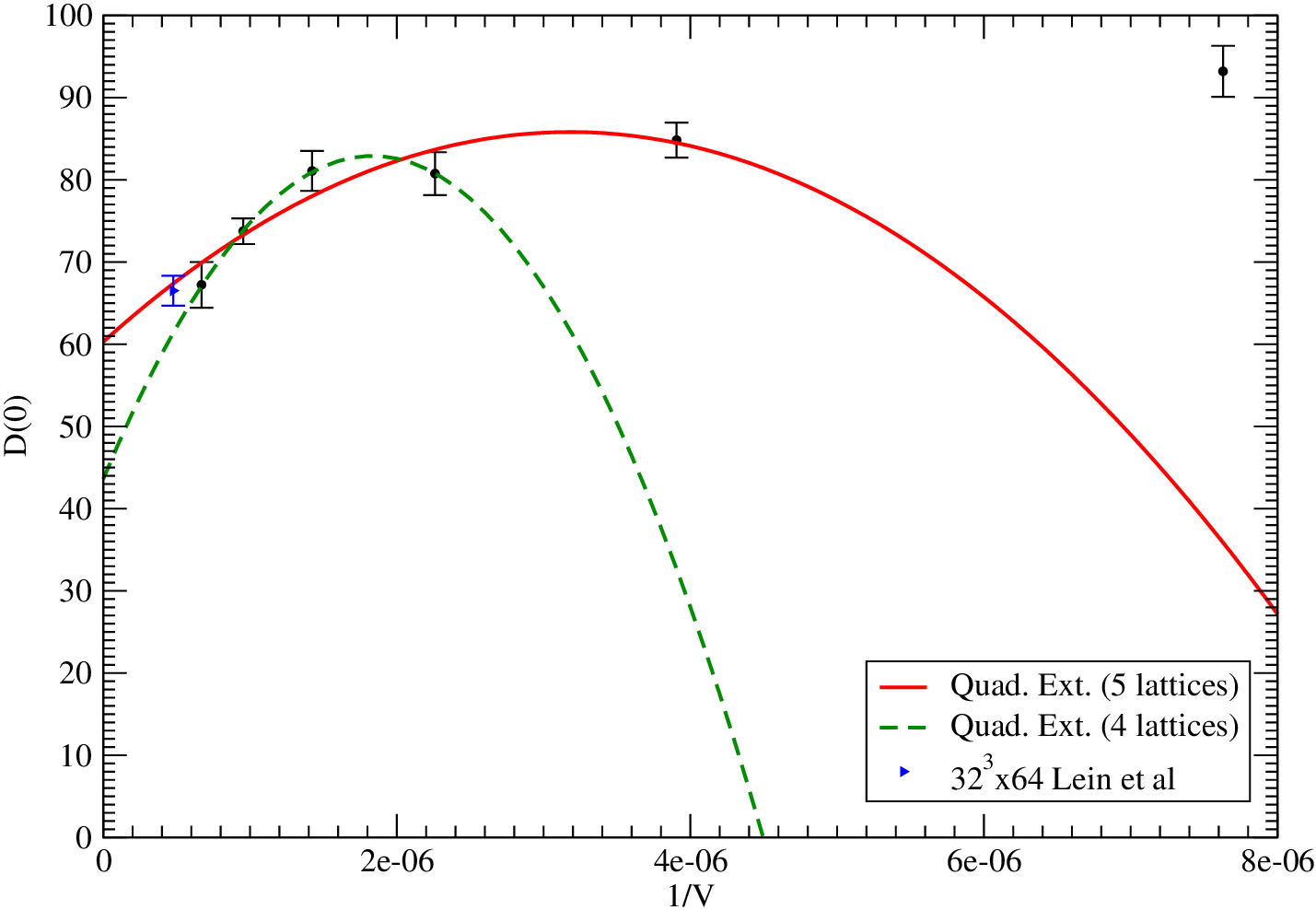}
\end{minipage}

\vspace{1.2cm}

\begin{minipage}[b]{\textwidth}
\centering
\includegraphics[origin=c,scale=0.5]{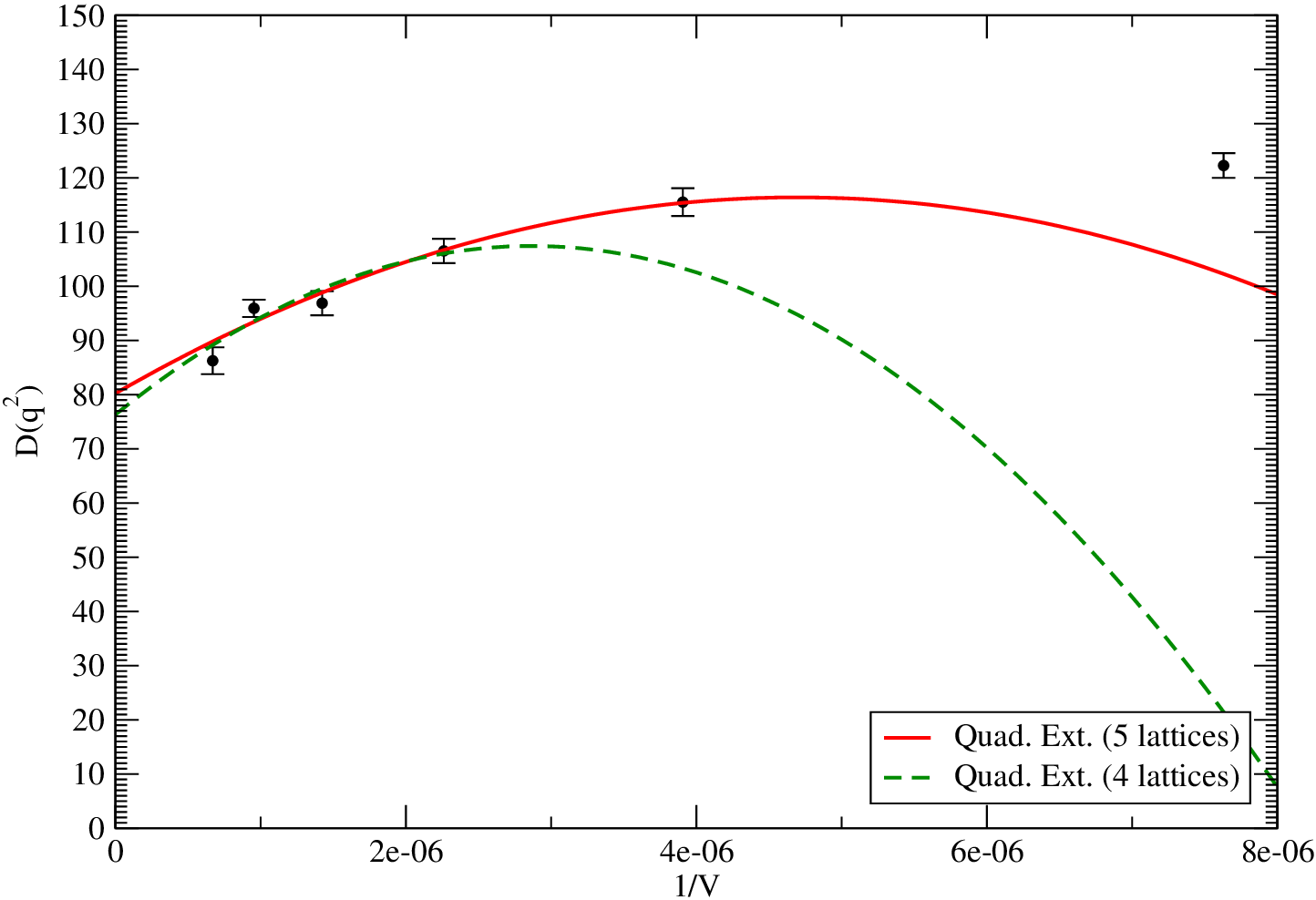}
\end{minipage}
\caption{Bare gluon propagator for the two smallest momenta as a function of 
the inverse volume. For comparison, for zero momentum we include $D(0)$ 
from the largest lattice of \cite{Lei99}, i.e. $32^3 \times 64$.}
\label{D01V}
\end{figure}

\begin{figure}[t]
\begin{minipage}[b]{\textwidth}
\centering
\includegraphics[origin=c,scale=0.5]{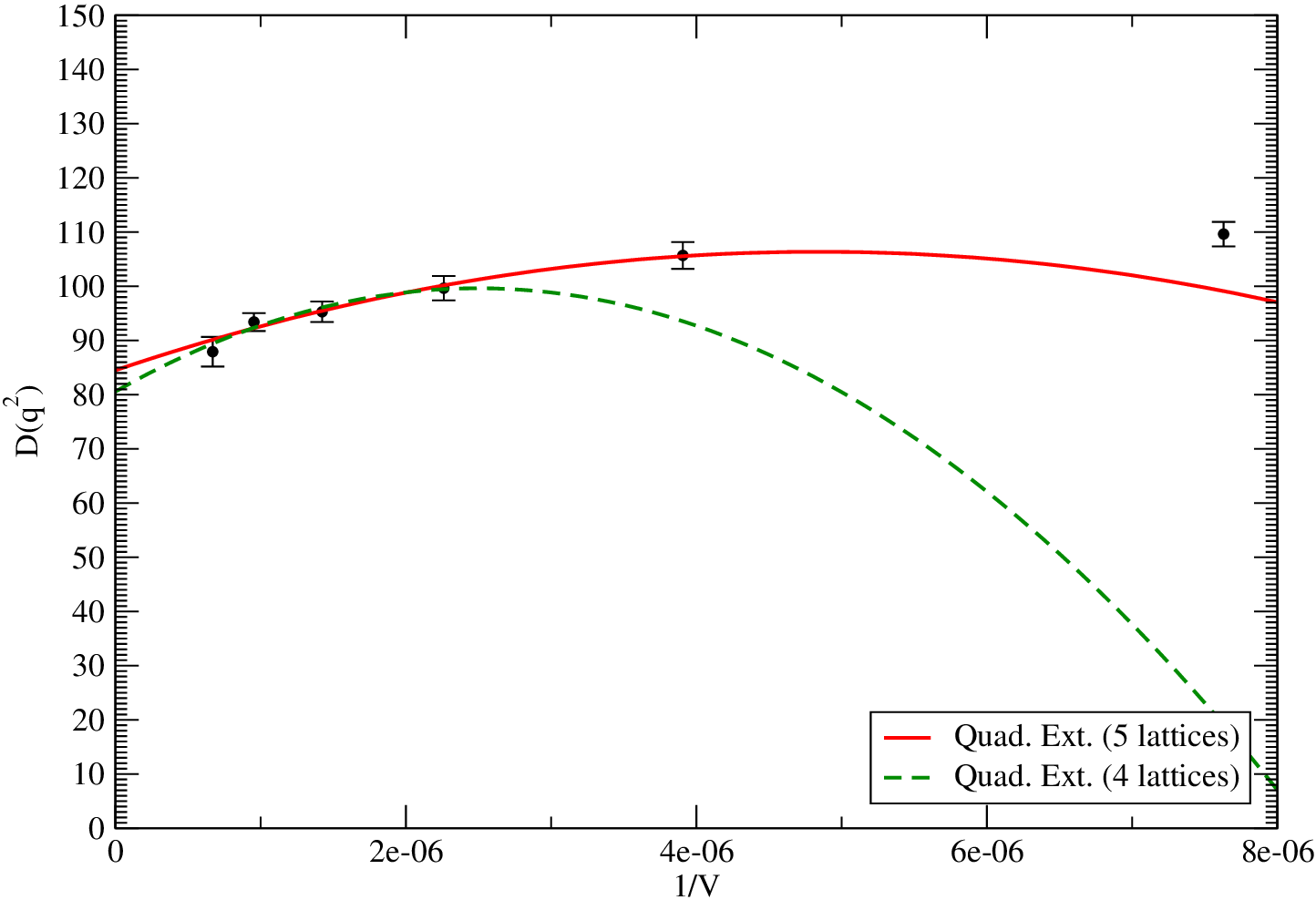}
\end{minipage}

\vspace{1.2cm}

\begin{minipage}[b]{\textwidth}
\centering
\includegraphics[origin=c,scale=0.5]{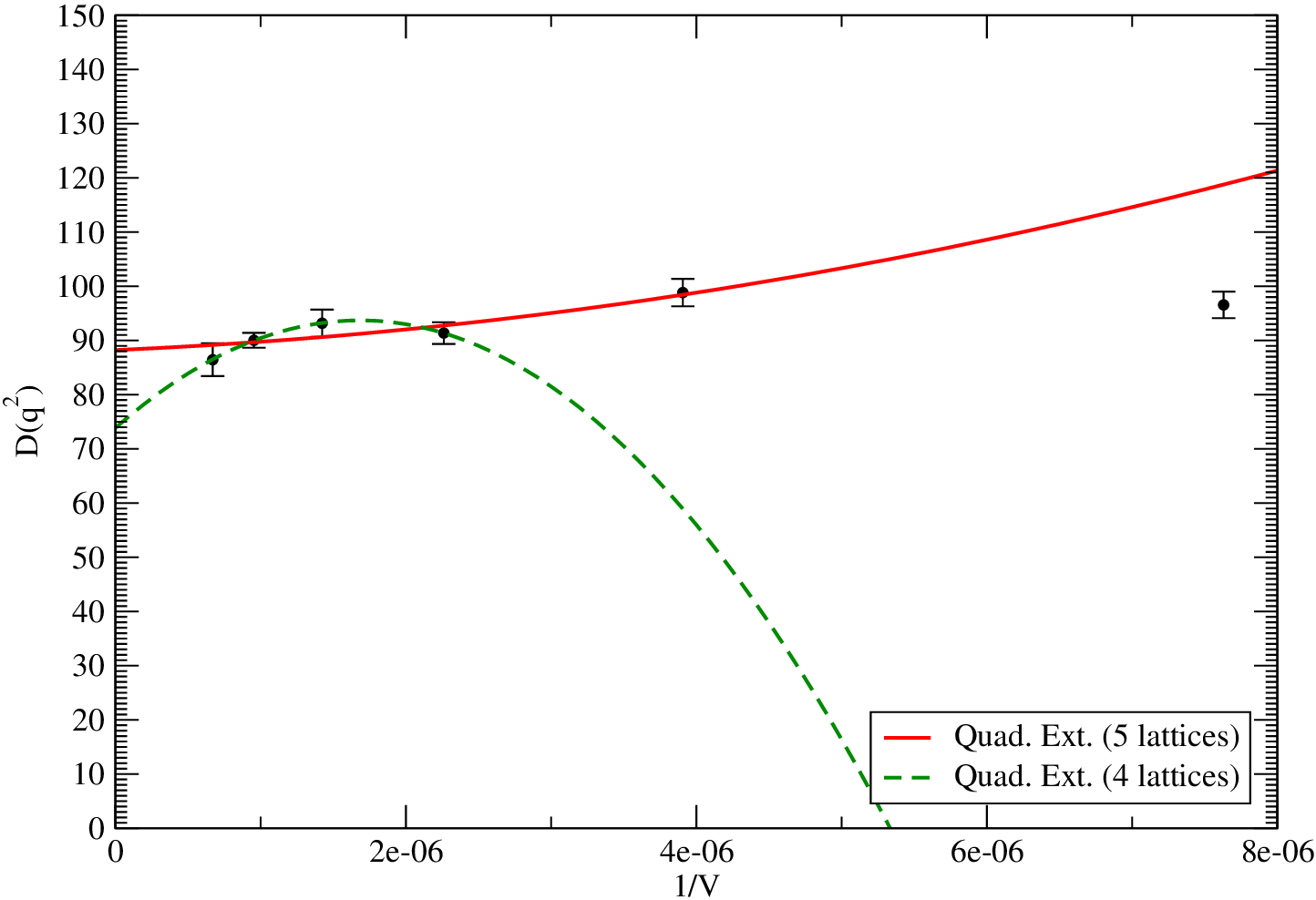}
\end{minipage} 
\caption{Bare gluon propagator as a function of the inverse volume for the
third and fourth smallest momenta.}
\label{D23V}
\end{figure}

So far we have extrapolated the $\kappa$ values. Alternatively, one can
extrapolate directly the gluon propagator and fit the associated dressing
function. In figures \ref{D01V} and \ref{D23V} the bare propagator is plotted,
as function of the volume, for the lowest four momenta. For all these momenta,
we tried linear, quadratic, cubic and quartic extrapolations. 
In table \ref{ChiVinfty} the $\chi^2/d.o.f.$ is reported for the 
various extrapolations considered and for the lowest four momenta. 

\begin{table}[t]
\begin{center}
\begin{tabular}{|r||rrrr|rrr|rr|}
\hline
   $p$          & \multicolumn{4}{c|}{Set I}
                & \multicolumn{3}{c|}{Set II}
                & \multicolumn{2}{c|}{Set III} \\
                & \multicolumn{4}{c|}{$8^3 - 18^3 \times 256$}
                & \multicolumn{3}{c|}{$10^3 - 18^3 \times 256$}
                & \multicolumn{2}{c|}{$12^3 - 18^3 \times 256$} \\
                & Lin & Quad & Cub & Quart &
                  Lin & Quad & Cub &
                  Lin & Quad \\
\hline
   0    & 2.84 & 2.60 & 1.54 & 0.21 &
          3.08 & 1.94 & 0.13 &
          2.99 & 0.01 \\
\hline
   1    & 6.15 & 1.95 & 2.60 & 4.53 &
          2.66 & 2.67 & 4.71 &
          2.92 & 5.07 \\
\hline
   2    & 2.02 & 0.44 & 0.56 & 0.71 &
          0.68 & 0.59 & 0.77 &
          0.72 & 0.90 \\
\hline
   3    & 1.66 & 1.02 & 1.32 & 0.06 &
          0.85 & 1.20 & 0.03 &
          1.04 & 0.00 \\
\hline
\end{tabular}
\caption{The $\chi^2/d.o.f.$ for the extrapolations to the infinite volume of 
the four lowest momenta gluon propagator assuming a linear (Lin), quadratic 
(Quad), cubic (Cub) and quartic (Quart) polinomial dependence on $1/V$ and for
various sets of data.} 
\label{ChiVinfty}
\end{center}
\end{table}

In the following, we will not consider extrapolations using all lattices 
(set I), because it involves our smallest lattice, which is too short 
$\sim 0.8$ fm in the spatial directions. Note, 
however, that the data reported in table \ref{ChiVinfty} point towards a
smooth approach to the infinite volume limit, starting from lattices sizes
as small as 0.8 fm.
Furthermore, figures \ref{D01V} and \ref{D23V} suggest that we should not 
use a linear extrapolation. According to the figures, a quadratic 
extrapolation, or a higher power of the inverse volume,
seems to be more suitable. However, for set II, and for the 
first non-vanishing momentum, the cubic extrapolation  gives a quite poor 
description of the data, when compared to the quadratic fit. So, for the 
reasons explained above, from now on we will consider only the quadratic 
extrapolation using set II and III.

Figures \ref{D01V} and \ref{D23V} show the lattice data together
with the two quadratic extrapolations to the infinite volume. Note that in
both extrapolations, the worst $\chi^2/d.o.f.$ is associated with the first 
nonzero momentum. Indeed, as can be seen in figure \ref{D01V}, for this 
momentum the data seems to fluctuate more than for the other momenta. For 
larger momenta the behaviour of the gluon propagator as a function of the 
volume becomes smoother.

In \cite{Bo01} the infinite volume of the renormalized zero momentum gluon
propagator was computed. The chosen renormalization condition was
\begin{equation}
    \left. D_R ( q^2 ) \right|_{q^2 = \mu^2} ~ = ~  \frac{1}{\mu^2} \, ;
\end{equation}
the lattice data was renormalized at $\mu = 4$ GeV. The quoted 
extrapolated renormalized zero momentum gluon propagator being 
$7.95(13)$ GeV$^{-2}$.
Our largest momentum available is $q = 3.88$ GeV, a slightly lower value.
Performing the renormalization in the same way but at the scale
$\mu = 3.88$ GeV, it comes
\begin{equation}
   D_R ( 0 ) ~ = ~ \left\{ \begin{array}{rcl}
             6.3 \pm 1.4 ~ \mbox{GeV}^{-2} & & 
                                  \mbox{set III data,} \\
            10.9 \pm 0.8 ~ \mbox{GeV}^{-2} & & 
                                  \mbox{set II data.}
                \end{array} \right.
\label{DR0}
\end{equation}
The quoted errors are pure statistical.
The value quoted in \cite{Bo01} is in between the above figures, with the 
result from the so called set III being almost compatible within one 
standard deviation with $7.95(13)$ GeV$^{-2}$ and the result from 
set II being compatible only at three standard deviations.

\begin{figure}[b]
\begin{minipage}[b]{\textwidth}
\centering
\includegraphics[scale=0.47]{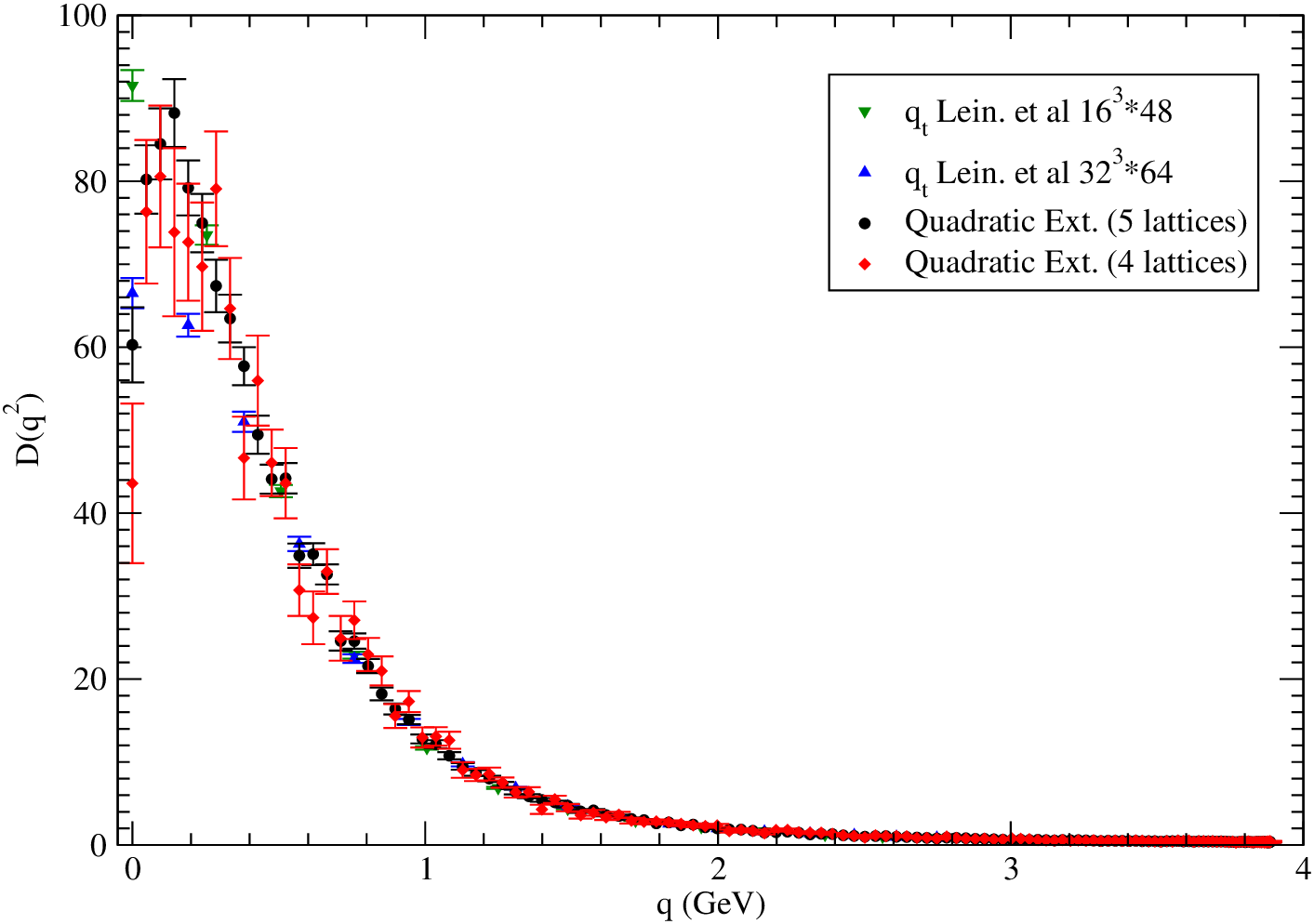}
\end{minipage}

\vspace{1.4cm}

\begin{minipage}[b]{\textwidth}
\centering
\includegraphics[scale=0.47]{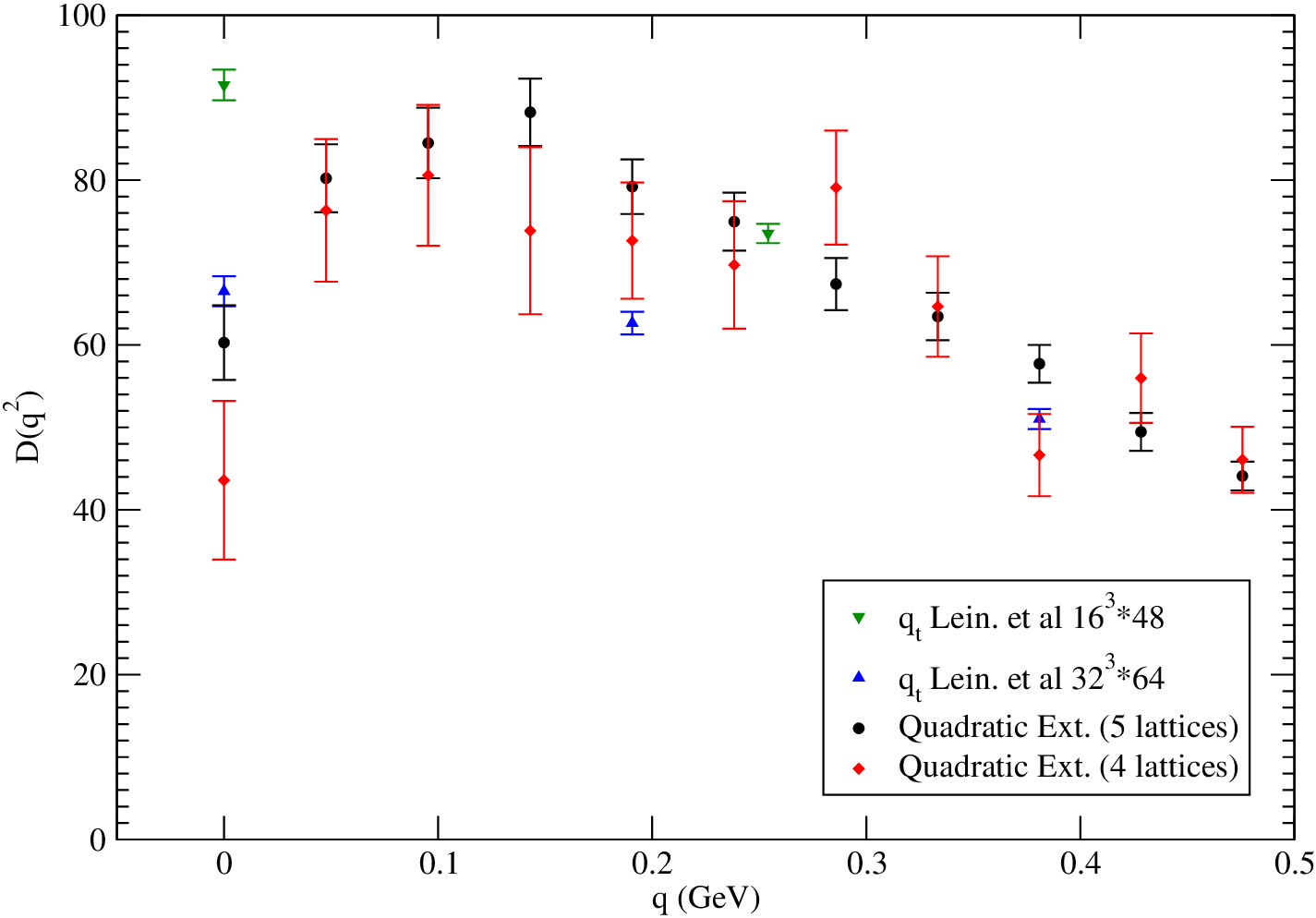}
\end{minipage} 
\caption{The bare extrapolated gluon propagator. 
The full range of temporal momenta is seen on the 
top with a zoom for the infrared region on the bottom. The errors on the 
propagator were computed assuming a gaussian error propagation. For 
comparisation, the figure includes the $16^3 \times 48$ and $32^3 \times 64$
propagators computed at $\beta = 6.0$ in \cite{Lei99}.}
\label{DExt}
\end{figure}

The extrapolated propagators can be seen in figure \ref{DExt}; the errors 
were computed assuming gaussian error propagation. 
For infrared momenta, the extrapolated propagators interpolate
between the two lattices simulated in \cite{Lei99}. Moreover, for zero momentum
the extrapolated propagators are smaller than the propagator of 
Leinweber \textit{et al} computed with their largest lattice.
The extrapolated 
propagators from fitting the two different sets of lattices are compatible, 
at least, at the level of two standard deviations. The propagator computed
from set III shows much larger statistical errors. The
differences between the two propagators are larger for smaller momenta.
This difference can be used to estimate systematic errors.

The fit of the dressing function computed from the extrapolated propagator, 
obtained from set II, to the first three nonzero 
momenta to the pure power law gives $\kappa = 0.5215(29)$ and has a 
$\chi^2/d.o.f. = 0.02$. Similar fits but using the corrections to (\ref{Zdse})
produce larger values for the same exponent ($\sim 0.55$). All the computed
$\kappa$'s are compatible with a vanishing zero momentum gluon propagator.
Moreover, the fit to the pure power law produces a number which agrees well 
with the  estimation from an extrapolation on $\kappa$ - see equations
(\ref{kinfty}) and (\ref{kinfty2}). If, instead, we perform the same analysis 
but for the extrapolated propagator obtained from set III,
then
$\kappa = 0.4979(66)$ and $\chi^2 / d.o.f. = 0.27$. Again, the fits with 
corrections to the pure power law give large values for 
$\kappa = 0.52 - 0.53$. Note that now the $\kappa$ fitting the pure power law 
is smaller than the corresponding value obtained from our largest lattice 
(see table \ref{kappa}) and is compatible with $\kappa = 0.5215(29)$ within 
two and a half standard deviations.

\section{Results and Conclusion}

We have computed the Landau gauge gluon propagator for large asymmetric 
lattices with different volumes. For each volume, we have checked that the 
lattice dressing function is well described by the DSE solution (\ref{Zdse}) 
for momenta below $\sim 150$ MeV by fitting the above functional form.
The exception being our smallest lattice $8^3 \times 256$.
Note that in all these fits $Z(q^2)|_{q=0}$ was never used.
In what concerns the infrared exponent $\kappa$, the data shows finite volume 
effects, with $\kappa$ becoming larger for larger volumes. In this sense, all 
values in table \ref{kappa} can be read as lower bounds on the infinite volume
figure. 

The infinite volume $\kappa$ was estimated in two different ways: i) from
an extrapolation of the values reported in table \ref{kappa}, assuming a
linear and a quadratic dependence on the inverse volume; ii) extrapolating the
propagator, assuming a quadratic dependence on $1/V$. The results of the first
method are (\ref{kinfty}), (\ref{kinfty2}), giving an weighted mean value of 
$\kappa = 0.5246(46)$. The value for the exponent from extrapolating directly 
the gluon propagator being $\kappa = 0.5215(29)$, if one considers the five
largest lattices (set II), and $\kappa = 0.4979(66)$, if one considers the 
four largest lattices (set III). 
The first value is on the top of the value obtained from 
extrapolating directly $\kappa$ as function of the volume. Moreover, these two 
values agree well with the time-independent stochastic quantisation prediction
$\kappa = 0.52145$ \cite{Zw02,Zw03} and 
are within the range of possible values 
estimated with renormalization group arguments \cite{Pa04,LiPa04,FiGi04},
$0.52 \, \le \, \kappa \, \le \, 0.595$. All 
these results point towards a vanishing zero momentum gluon propagator.
The $\kappa = 0.4979(66)$, obtained from fitting the gluon dressing function, 
computed from the quadratic extrapolation of the gluon propagator to the 
infinite volume, using only the four largest lattices (set III), 
is compatible with both
an infinite and a null zero momentum gluon propagator. This value is smaller
than the measured exponent from our largest lattice, $\kappa = 0.5017(49)$, 
and, although, these two numbers being compatible within errors, the 
extrapolated $\kappa$ does not follow the observed behaviour that $\kappa$
increases with the lattice volume. This is probably due to the large 
statistical errors observed in the extrapolation using the smaller set of 
lattices. Finally, one can claim a $\kappa = 0.498 - 0.525$ with the lattice 
data favouring the right hand side of the interval.

Although the lattice data seems to favour a null zero momentum gluon 
propagator, the measured extrapolated zero momentum propagator clearly does 
not vanish - see equation (\ref{DR0}). As a function of the lattice volume, 
the bare zero momentum propagator becomes smaller for larger volumes. 
Moreover, in \cite{NosDub2} it was shown that the extrapolation of the zero 
momentum propagator, depending on the functional form considered, is 
compatible with both a vanishing and a nonvanishing value. In this study we 
avoided the question of the zero momentum value by fitting only the smallest 
nonzero momenta. Solutions to this puzzle require simulations on larger 
lattices and/or having a better theoretical control on the volume 
extrapolation. Currently, we are engaged in improving the statistics for our
larger lattices and simulating bigger volumes, aiming to improve the infinite 
volume extrapolations.

\section*{Acknowledgements}

We would like to thank C. S. Fischer, J. I. Skullerud, A. G. Williams,
P. Bowman, D. Leinweber and A. C. Aguilar for inspiring discussions.
P. J. Silva acknowledges financial support from FCT via grant 
SFRH/BD/10740/2002. 


\end{document}